\documentclass[aps, prl, 10pt, twocolumn, superscriptaddress, noshowpacs, preprintnumbers, longbibliography, bibnotes,hyperref,floatfix]{revtex4-1}

\usepackage[colorlinks=true,breaklinks=true]{hyperref}
\hypersetup{allcolors=[rgb]{0.0 0.0 1.0},linkcolor=[rgb]{0.75 0.05 0.05}}
\usepackage[dvipsnames]{xcolor}
\usepackage{mathtools}
\usepackage{amsmath}
\usepackage{amsfonts}
\usepackage{amssymb}
\usepackage{bbold}
\usepackage{multirow}
\usepackage{bm}
\usepackage{hyperref}
\long\def\exclude#1{}
\usepackage{orcidlink}
\usepackage{slashed}
\usepackage[normalem]{ulem}
\usepackage[T1]{fontenc}

\DeclareRobustCommand{\okina}{%
  \raisebox{\dimexpr\fontcharht\font`A-\height}{%
    \scalebox{0.8}{`}%
  }%
}

\begin{document}

\title{The Black Hole Mass Gap as a New Probe of Millicharged Particles}

\author{Damiano F. G. Fiorillo \orcidlink{0000-0003-4927-9850}}
\email{damianofg@gmail.com}
\affiliation{Istituto Nazionale di Fisica Nucleare (INFN), Sezione di Napoli, Complesso Universitario di Monte Sant'Angelo, Via Cintia, 80126 Napoli, Italy}
\affiliation{Deutsches Elektronen-Synchrotron DESY,
Platanenallee 6, 15738 Zeuthen, Germany}

\author{Giuseppe Lucente
\orcidlink{0000-0003-1530-4851}}
\email{lucenteg@slac.stanford.edu}
\affiliation{SLAC National Accelerator Laboratory, 2575 Sand Hill Rd, Menlo Park, CA 94025}
\author{Jeremy Sakstein\,\orcidlink{0000-0002-9780-0922}} \email{sakstein@hawaii.edu}
\affiliation{Department of Physics \& Astronomy, University of Hawai\okina i, Watanabe Hall, 2505 Correa Road, Honolulu, HI, 96822, USA}

\author{\\ Edoardo Vitagliano
\orcidlink{0000-0001-7847-1281}}
\email{edoardo.vitagliano@unipd.it}
\affiliation{Dipartimento di Fisica e Astronomia, Università degli Studi di Padova,
Via Marzolo 8, 35131 Padova, Italy}
\affiliation{Istituto Nazionale di Fisica Nucleare (INFN), Sezione di Padova,
Via Marzolo 8, 35131 Padova, Italy}

\author{Matteo Cantiello
\orcidlink{0000-0002-8171-8596}}
\email{mcantiello@flatironinstitute.org}
\affiliation{Center for Computational Astrophysics, Flatiron Institute, 162 5th Avenue, New York, NY 10010, USA}
\affiliation{Department of Astrophysical Sciences, Princeton University, Princeton, NJ 08544, USA}

\begin{abstract}
{We investigate the impact of millicharged particles (MCPs) on massive stars undergoing pulsational pair-instability supernovae and on the location of the lower edge of the black hole mass gap. We find that energy losses due to MCP emission weaken the pulsations, allowing the star to retain more mass and thereby shifting the lower edge of the mass gap to higher black hole masses. The mass gap is sensitive to a region of MCP parameter space with masses $35\,{\rm keV}\lesssim m_\chi \lesssim 200\,{\rm keV}$ and charges $10^{-10}\lesssim q \lesssim 10^{-9}$, which remains unconstrained by existing astrophysical probes.
If confirmed, recent gravitational wave observations placing the lower edge of the mass gap near $45\,{\rm M}_\odot$ would translate directly into bounds on this parameter space.
}
\end{abstract}

\maketitle

\textbf{\textit{Introduction.---}}The Nobel prize-winning direct detection of gravitational waves (GWs) from GW150914 marked the birth of gravitational wave astronomy~\cite{LIGOScientific:2016vlm}.~Since then, over one hundred mergers have been observed~\cite{LIGOScientific:2018mvr,LIGOScientific:2020ibl,KAGRA:2021vkt,LIGOScientific:2025slb}, which have been used as a testbed for searches of new physics, particularly in the gravitational sector~\cite{Straight:2020zke,LIGOScientific:2019fpa,LIGOScientific:2020tif,LIGOScientific:2021sio}. Non-gravitational extensions of the Standard Model (SM), such as novel feebly interacting particles (FIPs), are harder to test through GWs alone. A multimessenger signal like GW170817~\cite{LIGOScientific:2017ync,LIGOScientific:2017zic,Burns:2019byj}, spanning GWs and electromagnetic radiation, may circumvent this challenge and provide competitive constraints (see, e.g.,~\cite{Dietrich:2019shr,Diamond:2021ekg,Fiorillo:2022piv,Sigurdarson:2022mcm,Diamond:2023cto,Dev:2023hax,Manzari:2024jns,Fiorillo:2025gnd}).~A compelling question remains, though:~can GWs from black holes (BHs) alone indirectly probe the existence of FIPs in regions not yet touched by other probes?

An interesting possibility in this context is the putative enhanced cooling induced by FIP emission on the formation of stellar mass BHs~\cite{Croon:2020ehi,Croon:2020oga,Sakstein:2020axg,Sakstein:2022tby}. This shows up most prominently in the lower edge of the so-called black hole mass gap (BHMG). This gap corresponds to the observed scarcity of BHs in the mass range $45-130\,{\rm M}_\odot$, corresponding to a range of initial stellar masses $130-250\,{\rm M}_\odot$, as the final BH mass can differ substantially from the initial mass due to the pair-instability. The gap, which was predicted long before gravitational-wave observations, arises from the production of $e^+e^-$ pairs after core-helium burning, which soften the equation of state and initiate a runaway collapse~\cite{Barkat:1967zz,1968Ap&SS...2...96F,1967ApJ...148..803R,2002RvMP...74.1015W,Belczynski:2016jno,2017ApJ...836..244W,2019ApJ...878...49W}. Contraction raises the temperature and density, producing more pairs and accelerating the collapse, terminating in the explosive ignition of oxygen, which either disrupts completely the star---a pair-instability supernova (PISN) with no remnant---or, for lighter progenitors, initiates a sequence of pulsations---a pulsational pair-instability supernova (PPISN)---which eject large amounts of mass, leaving a lighter BH. The gap terminates at higher masses~\cite{2002ApJ...567..532H}, above $120\,{\rm M}_\odot$,  due to photodisintegration preventing disruption, so that the star collapses directly into a black hole.~The existence of the gap in GW data remains an active area of research.~There was evidence for its existence in the LIGO/Virgo's second gravitational wave transient catalog (GWTC-2) \cite{Fishbach:2017zga}, but no conclusive signal was found in the larger LIGO/Virgo/KAGRA (LVK) GWTC-3 data set \cite{KAGRA:2021duu}.~Its presence in the most recent GWTC-4 catalog is debated \cite{Tong:2025wpz,Ray:2025xti,Antonini:2025ilj,Wang:2025nhf}.

\begin{figure*}
    \centering
    \includegraphics[width=1.\textwidth]{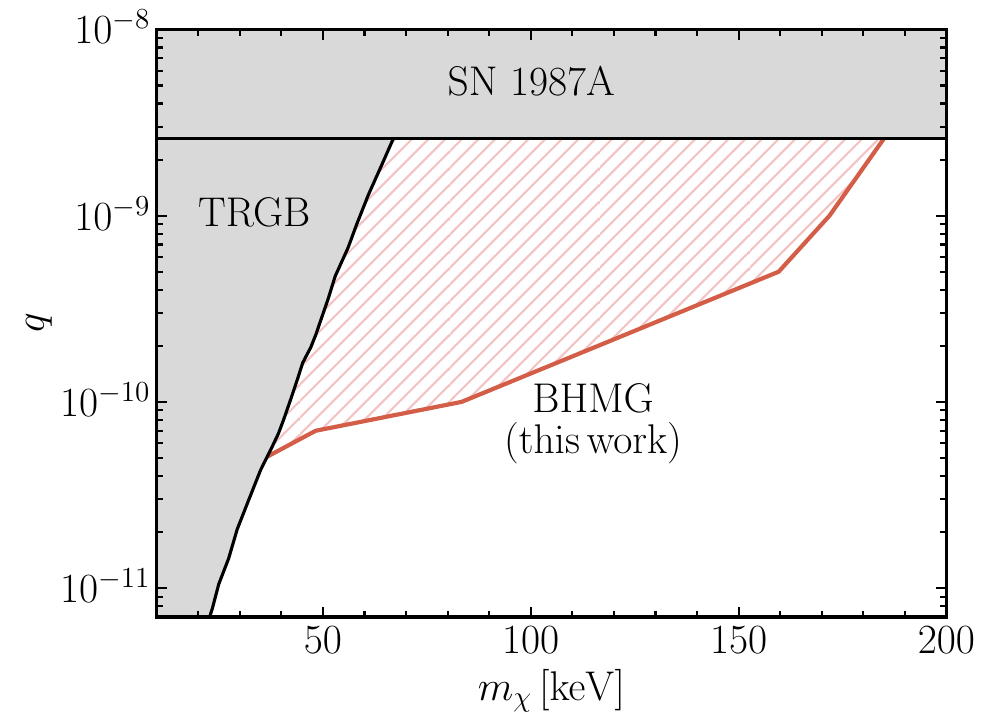}
    \caption{Sensitivity of the Black Hole Mass Gap (BHMG) to the fractional charge $q$ as a function of the MCP mass $m_\chi$. The hatched region will be excluded if the location of the lower edge of the BHMG is confirmed to be $45^{+5}_{-4}\,{\rm M}_\odot$ ($90\%$ credibility), as inferred from a recent analysis of the LVK GWTC-4 catalog~\cite{Tong:2025wpz}. Previous constraints from the cooling of SN~1987A~\cite{Fiorillo:2024upk} and the tip of the red-giant branch (TRGB)~\cite{Fung:2023euv} are shown in gray.}
    \label{fig:bounds}
\end{figure*}

The edges of the gap are known to be sensitive to the rate of the $^{12}{\rm C}(\alpha,\gamma)^{16}{\rm O}$ reaction \cite{Takahashi:2018kkb,2019ApJ...887...53F,Farmer:2020xne,Mehta:2021fgz,Farag:2022jcc,Croon:2023kct,Croon:2025gol,Xin:2026zyt}, whose experimental uncertainty propagates directly into uncertainty in the predicted gap location.~This rate remains highly uncertain due to experimental discrepancies \cite{deBoer:2017ldl,Shen:2023rco}, so it is standard to consider $\pm3\sigma$ variations around the median of current measurements \cite{2016MNRAS.456.3866C,deBoer:2017ldl,2019ApJ...887...53F,Farmer:2020xne,Mehta:2021fgz,Farag:2022jcc,2022ApJ...935...21C,2023ApJ...954...51C,Croon:2023kct}. The location of the BHMG is robust to variations in other stellar physics such as metallicity, winds, and other nuclear reaction rates \cite{2019ApJ...887...53F,Farmer:2020xne}.~Rotation is generally expected to play a subdominant role due to efficient angular momentum transport (e.g., via the Spruit–Tayler dynamo \cite{Fuller:2019sxi,Fuller:2022ysb}), although rapidly-rotating progenitors from channels such as chemically homogeneous evolution or tidal spin-up may contribute to a minority of systems \cite{deMink:2016vkw,Marchant:2016wow,Qin:2018vaa,Olejak:2021iux,Popa:2025dpz,Croon:2025gol}.

In addition to known stellar physics, the emission of light FIPs can also affect the location of the BHMG. Acting as additional stellar energy loss channels, FIPs enhance cooling in the stellar core and shorten the duration of core helium burning. As a consequence, the $^{12}{\rm C}(\alpha,\gamma)^{16}{\rm O}$ reaction has less time to convert carbon into oxygen, leading to a smaller $^{16}{\rm O}/^{12}{\rm C}$ ratio when the star enters the $e^+e^-$ pair-production regime. With less oxygen available to power the explosion and a larger convective carbon-burning shell resisting collapse \cite{2019ApJ...887...53F,Farmer:2020xne}, some stars that would otherwise undergo PPISN  avoid this fate, while those that do experience weaker pulses and eject less mass. As a result, the PPISN-PISN transition shifts to higher initial masses~\cite{Croon:2020ehi,Croon:2020oga,Sakstein:2020axg}.

This proposal has focused on FIPs lighter than the keV scale, which are efficiently produced in cooler stellar cores where their emission is more easily constrained. Consequently, existing astrophysical observations already impose stringent bounds that rule out couplings capable of significantly impacting the BHMG~\cite{Croon:2020oga}. (See, e.g., Ref.~\cite{Arza:2026rsl} for a recent review on existing constraints on FIPs.)

Here we show for the first time that the BHMG offers optimal sensitivity to FIPs with masses in the $\mathcal{O}(10\,-\,100)\,{\rm keV}$ range. As a benchmark, we consider sub-MeV millicharged particles (MCPs) with electric charges many orders of magnitude smaller than that of the electron. These particles cannot decay into SM particles and act solely as additional energy-loss channels. Conventional astrophysical probes based on cooler stellar environments cannot efficiently produce such heavy particles, while supernovae, despite their higher core temperatures, are sensitive to significantly larger couplings. Figure~\ref{fig:bounds} identifies a region of mass $m_\chi$ and fractional charge $q$ currently unconstrained by existing astrophysical observations, which would be excluded if the lower edge of the BHMG is located at $45^{+5}_{-4}\,{\rm M}_\odot$ ($90\%$ credibility), as inferred from a recent analysis of the LVK GWTC-4 catalog~\cite{Tong:2025wpz}.

\textbf{\textit{Millicharged particles.---}}Millicharged particles are new fermions $\chi$
charged under a hidden $ U(1)_{\rm H}$ symmetry that couples them to a dark photon (DP) mixing kinetically with the SM photon~\cite{Galison:1983pa,Holdom:1986eq} (see also Refs.~\cite{Holdom:1985ag,Dienes:1996zr,Goodsell:2009xc,Albertus:2026fbe}),
\begin{equation}
    \mathcal{L}\supset-\frac{1}{4}F'_{\mu\nu}F'^{\mu\nu}-\frac{\epsilon}{2}F'^{\mu\nu}F_{\mu\nu}
    +\bar{\chi}(i\gamma^\mu\partial_\mu+g_\chi \gamma^\mu A'_\mu-m_\chi)\chi.
\end{equation}
The field redefinition $A_\mu'\rightarrow A_\mu'-\epsilon A_\mu$ allows one to bring the kinetic term into its canonical form, revealing that the MCP has a charge $g_\chi \epsilon=e q \ll e$,
\begin{equation}
\mathcal{L} \supset q e\,\overline{\chi} \gamma^\mu\chi\,A_\mu + \overline{\chi}(i\slashed{\partial}-m_\chi)\chi \, .
\end{equation}

MCPs are constrained by laboratory, cosmological, and astrophysical observations~\cite{Jones:1976xy,
Dobroliubov:1989mr,Mohapatra:1990vq,Davidson:1991si,Mitsui:1993ha,
Davidson:1993sj,Prinz:1998ua, Davidson:2000hf,Dubovsky:2003yn,Gies:2006ca,Gninenko:2006fi,
Badertscher:2006fm,
Jaeckel:2009dh,
Jaeckel:2010ni,Diamond:2013oda,
Vogel:2013raa,Moore:2014yba, Vinyoles:2015khy,Chang:2018rso,Magill:2018tbb,Gan:2023jbs}.  The dominant bounds for masses between keV and MeV come from the observations of supernova~(SN)~1987A and low-energy SNe~\cite{Fiorillo:2024upk} (see however Ref.~\cite{Fiorillo:2023frv} for a  comparison of modern neutrino cooling simulations and SN 1987A data), extending to masses of tens of MeV but only reaching down to $q_{\rm SN}\sim 10^{-9}$, and the luminosity of the tip of the red-giant branch (TRGB)~\cite{Fung:2023euv}, reaching down to $q_{\rm TRGB}\sim 10^{-14}$ but only extending to $m_\chi\lesssim 10\,\mathrm{keV}$. This limited mass reach comes from the relatively low temperature of the stellar cores, never surpassing $T\simeq 10^8 \, \rm K\simeq 8.6\rm\, keV$. On the other hand, late-stage evolution of pre-SNe massive stars can produce MCPs as heavy as $m_\chi\simeq \mathcal{O}(100\,\rm keV)$, a little lighter than the electron, as the temperature of the plasma becomes larger than $10\,\rm keV$ at the onset of the helium burning phase, and even hotter temperatures are reached during the burning of heavier elements. In fact, the very nature of PPISN,
based on the production of electron-positron pairs, immediately implies that novel particles lighter than electrons can be produced efficiently prior to the instability.

\begin{figure*}[t!]
    \centering \includegraphics[width=0.49\textwidth]{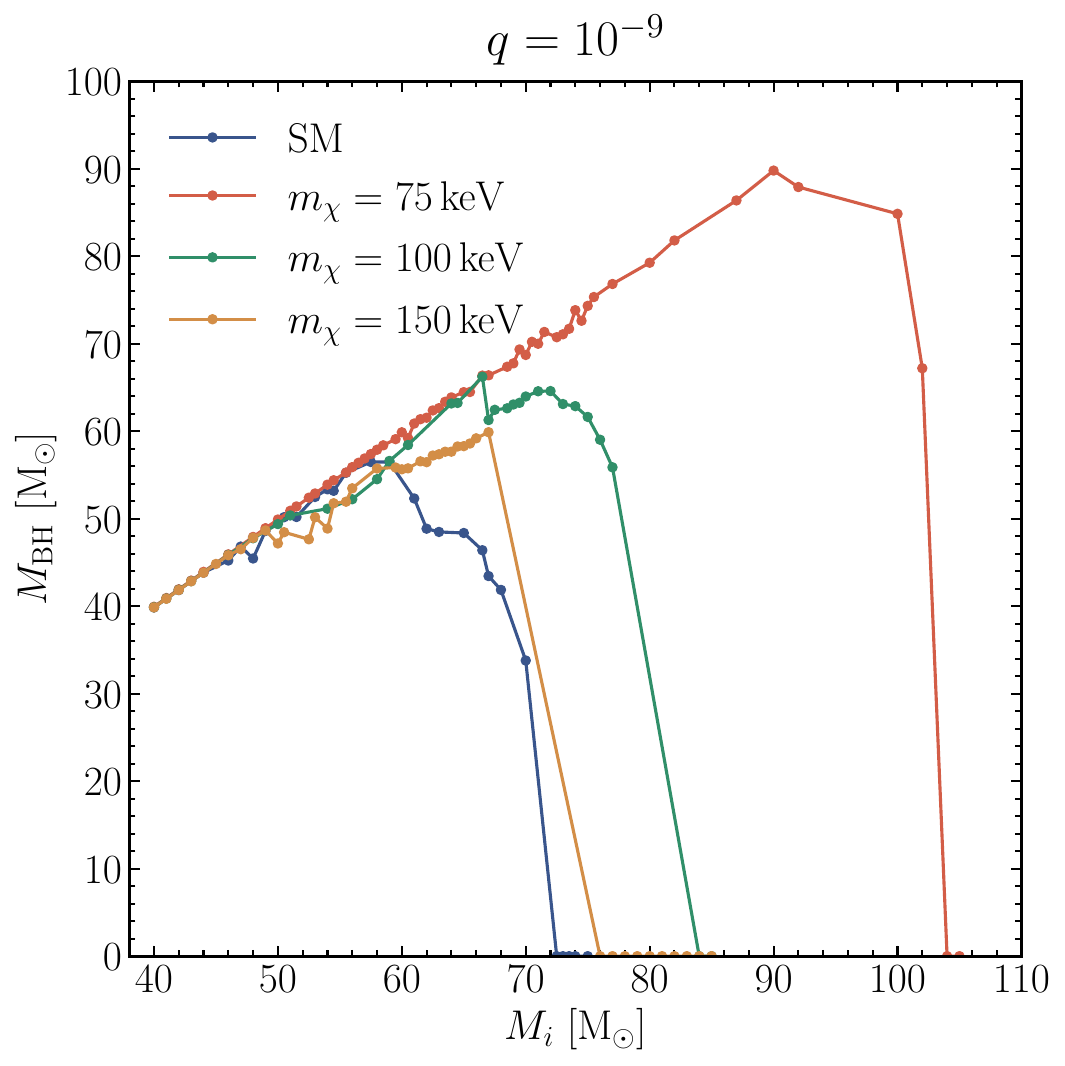}
\includegraphics[width=0.49\textwidth]{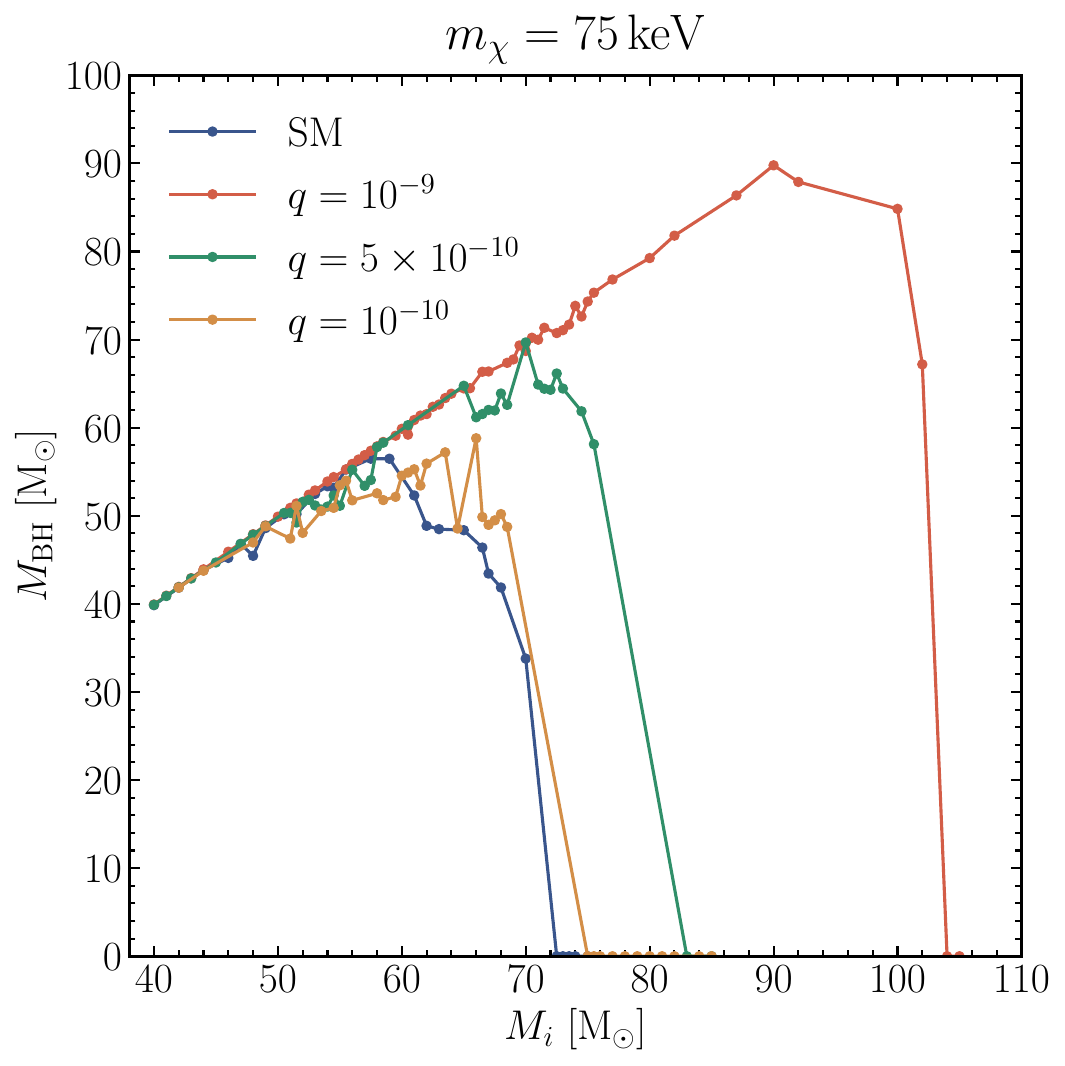}
    \caption{Black hole mass as a function of initial helium core mass for fixed $q=10^{-9}$ with varying MCP mass $m_\chi$ (left), and for fixed $m_\chi=75$ keV with varying $q$ (right). {Both panels were obtained assuming median $^{12}{\rm C}(\alpha,\gamma)^{16}{\rm O}$ rates.}
    }
    \label{fig:BHMS}
\end{figure*}

\textbf{\textit{Production rates in pre-supernovae.}}---Light MCPs in stellar cores are produced predominantly by the decay of plasmons, which is only kinematically allowed for MCPs roughly lighter than the plasma frequency $\omega_{\rm pl}\simeq 1\,{\rm keV}$ 
during the helium-burning phase. Heavier MCPs, which are the main target of this work, behave as electrons and positrons with a smaller charge, and are therefore produced by analogous processes, in particular Compton emission $e^-+\gamma \to e^-+\bar{\chi}+\chi$ and pair production $e^-+e^+\to \bar{\chi}+\chi$.

We have recently provided a comprehensive determination of the volumetric emissivity for all these processes~\cite{Prod}. For our purposes, we only require the energy-integrated cooling rate, for which simple analytical expressions are given in Ref.~\cite{Prod}. Here we summarize the dominant scalings with the density and temperature of the emitting plasma. The volumetric cooling rate $Q_\chi = d\mathcal{E}/dVdt$ for Compton emission scales as $Q_{\rm Com} \sim q^2 \alpha^3  n_e T^2 \mathrm{min}[1, T^2/m_e^2]e^{-m_\chi/T}$, where the exponential factor accounts for the Boltzmann suppression when $m_\chi\gtrsim T$. For pair production, the density dependence drops out since the product of electron and positron densities is independent of the chemical potential in non-degenerate conditions. The cooling rate therefore scales as $Q_{\rm pair} \sim q^2 \alpha^2 T^5$ for $T \gg m_e$, while at lower temperatures positrons are Boltzmann suppressed.

During core helium burning, the typical temperature is $T \sim 30\text{--}40\,\mathrm{keV}$, so MCP production is dominated by non-relativistic Compton scattering. As the temperature increases, $e^+e^-$ annihilation becomes increasingly important and dominates for $T \gtrsim m_e$. We therefore adopt the emissivities for these processes from Ref.~\cite{Prod} in our simulations, while neglecting plasmon decay due to the relatively low core densities during stellar burning. Including this additional channel would only strengthen the sensitivity of the BHMG.

As discussed above, millicharged particles affect the BHMG by shortening the duration of core helium burning. This suggests that the BHMG can probe MCPs when their energy-loss rate per unit mass, $\varepsilon_\chi = Q_\chi/\rho$, becomes comparable to neutrino losses at helium depletion (HD) \cite{Croon:2020oga}. At this stage, $T_{\rm HD}\simeq 3\times 10^{8}\,{\rm K}$, $\rho_{\rm HD}\simeq 10^{3}\,{\rm g/cm^3}$, and $\varepsilon_\nu^{\rm HD}\simeq10^3\,{\rm erg\,g^{-1}\,s^{-1}}$~\cite{Haft:1993jt,Raffelt:1996wa}, with MCP production dominated by Compton scattering.~To estimate if the BHMG probe is competitive with existing constraints, we compare with analogous conditions in other stellar environments. For the TRGB constraint, one requires $\varepsilon_\chi < \varepsilon_\nu^{\rm RG} \simeq 10\,{\rm erg\,g^{-1}\,s^{-1}}$, evaluated at $T_{\rm RG}=8.6\,{\rm keV}$ and $\rho_{\rm RG} \simeq 10^{6}\,{\rm g/cm^3}$, while for SN 1987A one finds $\varepsilon_\chi < \varepsilon_{\rm SN} \simeq 10^{19}\,{\rm erg\,g^{-1}\,s^{-1}}$ at $T_{\rm SN}=30\,{\rm MeV}$ and $\rho_{\rm SN} \simeq 3\times10^{14}\,{\rm g/cm^3}$~\cite{Raffelt:1996wa}.~In the low-mass limit, MCP production is dominated by plasmon decay, $Q_{\rm Dec}\sim q^2 \alpha^2 T^3 \omega_{\rm pl}^2$, so TRGB constraints dominate over the BHMG sensitivity in this regime. In contrast, matching the relative importance of MCP and neutrino cooling in the two environments---Compton emission at HD and plasmon decay in supernovae---suggests the BHMG can surpass SN bounds for MCP masses $m_\chi \lesssim 200\,{\rm keV}$.

\textbf{\textit{Simulations.}}---Our region of interest in Fig.~\ref{fig:bounds} was found by implementing the MCP pair-production rates described above into the stellar structure code MESA version 12778 \cite{Paxton:2010ji,Paxton:2013pj,Paxton:2015jva,Paxton:2017eie,Paxton:2019lxx,MESA:2022zpy}. The details of how (P)PISN are treated in \textsc{MESA} can be found in, e.g., Refs.~\cite{2019ApJ...882...36M,2019ApJ...887...53F,Croon:2020oga,Farmer:2020xne,Farag:2022jcc,Croon:2023kct,Croon:2025gol}.~The code includes all relevant physics such as Wolf-Rayet winds, which follow the prescription of \cite{Brott:2011ni}, overshooting, and departures from hydrostatic equilibrium, which are handled using a Harten-Lax-van Leer  solver \cite{Paxton:2017eie}.~Our modifications and a complete list of stellar physics adopted can be found in our reproduction package, which will be made available upon publication.~Here we highlight two important features: (1) we use the state-of-the-art $^{12}{\rm C}(\alpha,\gamma)^{16}{\rm O}$ reaction rate reported by \cite{Mehta:2021fgz}; and (2) we adopt the resolution controls {\tt delta\_lgRho\_cntr\_limit = 0.001d0} and {\tt max\_dq = 5d-4}, which are required to ensure that the core collapse--PPISN transition is accurately resolved~\cite{Mehta:2021fgz,Farag:2022jcc,Croon:2023kct}.

We simulated the evolution of non-rotating helium cores with metallicity $Z=10^{-5}$ to their final fate --- either core collapse (possibly via PPISN) or complete disruption due to PISN. The hydrogen envelopes of such massive stars are expected to be stripped by strong stellar winds or binary interactions, leaving behind bare helium cores that determine the subsequent evolution~\cite{Heger:2001cd,Eldridge:2007mi,Belczynski:2016obo,Woosley:2016hmi}. In Fig.~\ref{fig:BHMS} we show the final black hole mass, defined as the mass of bound material at core collapse, as a function of the initial helium core mass $M_i$. Without the inclusion of MCPs (blue line) stars with $M_i\lesssim 60\,{\rm M}_\odot$ directly collapse to black holes ($M_{\rm BH}\approx M_{i}$), for $60\,{\rm M}_\odot \lesssim M_i\lesssim 70\,{\rm M}_\odot$ stars undergo PPISN ($M_{\rm BH} < M_{i}$), while for larger masses stars experience PISN which completly disrupt them ($M_{\rm BH} = 0$). The lower edge of the BHMG is defined by the heaviest BH that can be formed, and Fig.~\ref{fig:BHMS} shows that this edge is shifted to higher BH masses in the presence of MCPs.

The trends in Fig.~\ref{fig:BHMS} reflect the cooling mechanism described above. At fixed MCP mass, increasing $q$ enhances the emission rate and shifts the lower edge of the gap to higher black hole masses. Conversely, at fixed $q$, the effect diminishes for larger $m_\chi$ due to Boltzmann suppression of the emission rate.

\begin{figure}[t!]
    \centering
    \includegraphics[width=0.49\textwidth]{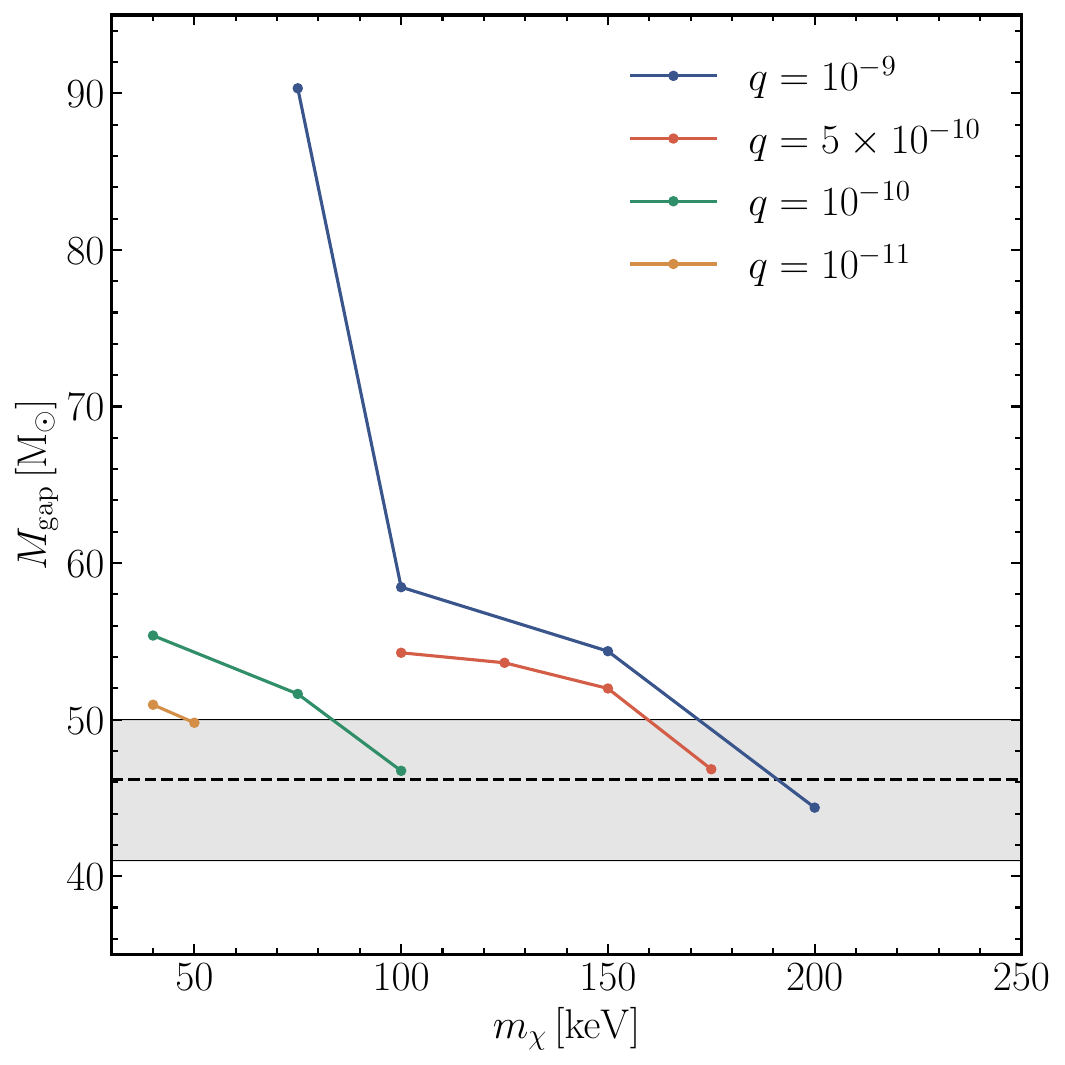}
    \caption{Lower edge of the black hole mass gap as a function of the MCP mass $m_\chi$ ($x$-axis) and $q$ assuming the $+3\sigma$ $^{12}{\rm C}(\alpha,\gamma)^{16}{\rm O}$ rate.~Each colored point corresponds to a different value of $q$ indicated in the figure.~The black dashed line is the Standard Model prediction,
    and the gray region indicates the 95\% C.L. reported by \cite{Tong:2025wpz}.~The hatched red region in Fig.~\ref{fig:bounds} was found by interpolating each colored curve and finding the value of $q$ for which it intersects the gray region.}
    \label{fig:mvsgap}
\end{figure}

The MCP parameter space can be probed by requiring that the shifted location of the lower edge of the BHMG induced by MCPs is consistent with observations. Recently, the LVK Collaboration released data from its fourth observing run (O4a)~\cite{LIGOScientific:2025slb}. Reference~\cite{Tong:2025wpz} reports a measurement of the black hole mass gap in this data at $45^{+5}_{-4}\,{\rm M}_\odot$ (95\% C.L.), obtained by fitting a parametric black hole mass function that includes a gap. Subsequently, Ref.~\cite{Ray:2025xti} reported no evidence for a mass gap when using a different model that does not allow for sharp features. The Bayes factor for this smooth model is larger by a factor of $\sim 3$, although its maximum likelihood is smaller by a factor of $\sim 2$. More data will be required to definitively establish the existence of a black hole mass gap; however, if the measurement reported in Ref.~\cite{Tong:2025wpz} is confirmed, {the hatched red region in the MCP parameter space shown in Fig.~\ref{fig:bounds} will immediately become a new exclusion zone.} 

The region shown in Fig.~\ref{fig:bounds}  was obtained as follows. We fixed the $^{12}{\rm C}(\alpha,\gamma)^{16}{\rm O}$ reaction rate to $+3\sigma$, its most extreme plausible value. With this choice, the SM prediction for the location of the mass gap is $46\,{\rm M}_\odot$, consistent with the reported measurement~\cite{Tong:2025wpz}. We then scanned the MCP parameter space to identify the smallest values of $\{m_\chi,q\}$ that remain compatible with the reported gap location. Because the parameter space must be sampled discretely, we interpolated our scans to avoid over-estimating the bounds. This procedure is illustrated in Fig.~\ref{fig:mvsgap} and is conservative:~the $+3\sigma$ $^{12}{\rm C}(\alpha,\gamma)^{16}{\rm O}$ rate yields the lowest plausible SM prediction for the gap location. Increasing this rate would shift the SM gap to higher black hole masses, making MCPs easier to exclude.~Statistically, the $+3\sigma$ $^{12}{\rm C}(\alpha,\gamma)^{16}{\rm O}$ rate corresponds to an extreme value within the allowed range. A full Bayesian treatment of the reaction rate uncertainty jointly with the MCP parameters would place most of the statistical weight at lower rates, for which the Standard Model prediction yields a higher gap location, and would therefore likely strengthen the resulting bounds. Finally, we note that our decision to omit rotation from our simulations is also conservative \cite{Marchant:2020haw,Croon:2025gol}, since rapidly rotating progenitors may lead to higher gap locations, thereby strengthening the resulting bounds on MCPs.

\textbf{\textit{Discussion and outlook.}}---Many extensions of the Standard Model of particle physics introduce a new $\rm U(1)_{H}$  gauge group with an associated massless boson, the dark photon, and fermions charged under $\rm U(1)_{H}$. If the dark photon kinetically mixes with the Standard Model photon, these fermions acquire a small effective electric charge and behave as millicharged particles (MCPs).~Millicharged particles  can be abundantly produced in the cores of massive stars, as their tiny electric charge allows efficient production without impeding their escape. We find that the lower edge of the black hole mass gap can probe unconstrained regions of the parameter space if putative MCPs have masses $35\,{\rm keV}\lesssim m_\chi \lesssim200\,{\rm keV}$ and charges $10^{-10}\lesssim q\lesssim10^{-9}$, demonstrating for the first time that the BHMG is sensitive to particles with masses above the keV scale. The BHMG thus provides a competitive probe of feebly interacting particles with masses slightly below the electron mass.

While we have focused on MCPs, the BHMG may also probe other models. Our results identify two conditions: (i) that the feebly interacting particle $X$ be light enough to be produced;~and (ii) to avoid stringent bounds from decays into visible SM particles, it must either be stable or have a small branching ratio into SM particles. The latter requirement excludes, for example, axion-like particles coupling to two photons from having a large impact on the BHMG, even in the keV-MeV mass range~\cite{Candon:2024eah}.

Finally, it is still possible that at least part of the region of interest identified in this paper could be excluded by different observations, such as the $R_2$-parameter, i.e., the ratio of stellar populations on the Asymptotic
Giant Branch to Horizontal Branch in Globular Clusters~\cite{2016MNRAS.456.3866C,Dolan:2022kul,Dolan:2023cjs}. We plan to explore this possibility in a forthcoming paper.

\section*{Acknowledgments} 
We thank Mathieu Renzo for helpful discussions at the early stages of this paper. We are grateful to Djuna Croon for helpful comments on the final manuscript. This article is based on work from COST Action COSMIC WISPers
(CA21106), supported by COST (European Cooperation
in Science and Technology).
D.F.G.F. acknowledges support by the TAsP (Theoretical Astroparticle Physics) project, and was supported by the Alexander von Humboldt Foundation (Germany) for most of the completion of the project.
G.L. acknowledges support from the U.S. Department of Energy under contract number DE-AC02-76SF00515.~J.S.~is supported by NSF Grant No.~2207880.~E.V. acknowledges support from the Italian Ministero dell'Università e della Rircerca through the FIS 2 project FIS-2023-01577 (DD n. 23314 10-12-2024, CUP C53C24001460001) and through Departments of Excellence grant 2023--2027 ``Quantum Frontier'', as well as from Istituto Nazionale di Fisica Nucleare (INFN) through the Theoretical Astroparticle Physics (TAsP) project. The Center for Computational Astrophysics at the Flatiron Institute is supported by the Simons Foundation.

\bibliographystyle{bibi}
\bibliography{References}

\end{document}